\documentclass[final,english]{bullsrsl}

\usepackage[latin1]{inputenc}

\usepackage[T1]{fontenc}

\usepackage{amsmath}
\usepackage{natbib} 

\usepackage{graphicx}

\begin{document}
\title{Feedback-Driven Star Formation and YSO Distribution around H{\sc~ii} Regions Sh2-140 and Sh2-145 }


  \author[affil={1,2}, corresponding]{C}{Rishi}
  \author[affil={1}]{Neelam}{Panwar}
  \author[affil={3}]{Ajay}{Kumar Singh}

\affiliation[1]{Aryabhatta Research Institute of Observational Sciences (ARIES), Manora Peak, Nainital 263002, India}
\affiliation[2]{Mahatma Jyotiba Phule Rohilkhand University, Bareilly, 243006, Uttar Pradesh, India}

\affiliation[3]{Department of Applied Physics/Physics, Bareilly College, Bareilly-243001, Uttar Pradesh, India}



\correspondance{rishictk@gmail.com}


\maketitle

\begin{abstract}
OB2 associations are ideal sites for studying various star-formation processes. Here, we study the region containing the bright-rimmed cloud (BRC)~44 and the dark cloud LDN 1204, which lies between two H {\sc ii} regions, Sh2-140 and Sh2-145. The region is a part of a larger Cepheus OB2 association. Our previous study on BRC~44 suggests a scenario of triggered star formation through radiative feedback. Using the infrared data from the {\it Spitzer Space Telescope} and the Wide-field Infrared Survey Explorer, we find a spatial correlation between the locations of massive stars and the distribution of young stellar objects, indicating a possible sequential pattern of star formation across the region. These results indicate the role of massive-star feedback in shaping the star-formation history of the complex.
\end{abstract}

\keywords{H II regions (694); Low mass stars (2050); Star forming regions (1565); Star formation (1569)}


\section{Introduction}
Star-forming regions (SFRs) provide valuable opportunities to study the formation, evolution, and dynamics of molecular clouds and young stellar objects (YSOs). Detailed analysis of stellar populations in young, massive SFRs enables quantification of how the local environment influences star formation over time, the properties and distribution of ionized gas, and the dispersal and remnant structure of parent molecular clouds \citep{2010RSPTA.368..713L}. The presence of high-mass stars
in these environments significantly affects both ongoing and future star formation processes \citep{2012PhDT.........4T}. As massive stars form, they emit a substantial amount of ultraviolet (UV) radiation.
These photons ionize and heat the surrounding gas, resulting in the formation of bright H {\sc ii} (ionized hydrogen) regions and driving ionization and shock fronts into the surrounding molecular cloud. This radiative and mechanical feedback can either compress nearby gas to trigger subsequent star formation or disperse the cloud material, thereby suppressing further star formation \citep{2018BKAS...43Q..43K,2012MNRAS.420..562H}.
 Understanding this dual role of feedback is essential for developing a comprehensive picture of star formation in clustered environments.

The influence of massive-star feedback on subsequent star formation can be investigated through the spatial distribution and evolutionary status of YSOs within molecular clouds. Triggered star formation models, such as radiation-driven implosion (RDI; \citep{2011MNRAS.412.2079M}) and collect-and-collapse \citep{2008A&A...482..585D}, predict that newly formed stars should exhibit characteristic spatial patterns relative to ionization fronts, bright rims, and molecular cloud structures. In particular, younger protostars (Class 0/I sources) are often found embedded within dense cloud material, while relatively evolved Class II/ Class III sources are comparatively closer to the ionizing sources. The presence of age gradients, aligned YSO distributions, and correlations between stellar populations and cloud morphology can therefore provide important clues regarding the role of feedback in regulating star formation.

However, YSOs are often embedded within dense molecular clouds and are subject to substantial dust extinction; optical observations alone are often insufficient to identify the young stellar population. Infrared observations provide an effective alternative because circumstellar disks and envelopes surrounding YSOs absorb stellar radiation and re-emit it at longer wavelengths, producing characteristic infrared excess emission. Consequently, infrared surveys have become one of the most powerful tools for identifying and classifying YSOs and for investigating the early stages of star formation. Deeply embedded Class 0/I sources generally exhibit stronger infrared excesses than the more evolved Class II and Class III objects \citep{2010AJ....140.1214C}. Therefore, the spatial distribution of YSOs across different evolutionary classes serves as a valuable probe of the temporal and spatial evolution of star formation within molecular clouds \citep{2016A&A...586A..44C}.
Such spatial distribution, together with age gradients and cloud morphology, is often interpreted as evidence for sequential or feedback-triggered star formation, although additional age and kinematic information is generally required for confirmation.
 As a result, infrared surveys have become one of the most powerful tools for investigating the formation and early evolution of stellar populations \citep{2017MNRAS.468.2684P}.

In this work, we focus on a star-forming complex that is part of the larger OB association, Cepheus OB2. This region contains the bright-rimmed cloud (BRC) BRC~44 \citep{2008A&A...477..557M}, which is located at the northeastern periphery of the H\,{\sc ii} region Sh2-145 \citep{2025ApJ...994..222C}, as well as the dark cloud LDN~1204 \citep{2002AJ....124.2152B}, which is interacting with the two H\,{\sc ii} regions, Sh2-145 (distance $\sim$ 930 pc \citep{2015AJ....150..147F}) and Sh2-140 (distance $\sim$ 906 pc \citep{2023MNRAS.520.1390S}). For simplicity, we take the distance to the entire complex as $\sim$ 900 pc. The ionizing sources of the Sh2-145 are HD 213023 (O9V) and BD 62 2078 (O7.5V) \citep{1994A&A...283..963C} while, the Sh2-140 is excited by the B0.5\,V star HD~211880 \citep{1996A&A...315L.281T}. In our previous study of BRC~44, we found evidence suggesting that radiative feedback from the massive stars associated with Sh2-145 has influenced the evolution of the cloud and may have contributed to a subsequent episode of star formation within the bright rim, consistent with the RDI scenario \citep{2025ApJ...994..222C}. Motivated by these results, we extend the investigation to the wider region to examine the large-scale influence of radiation feedback on the surrounding molecular clouds. 

Using infrared observations from the \textit{Spitzer Space Telescope} and the Wide-field Infrared Survey Explorer (WISE), we identify and characterize the young stellar population across the complex. The primary goals of this study are to examine the spatial distribution of YSOs and assess the role of massive-star feedback in shaping the star-formation history of the complex. The paper is organized as follows. In Section 2, we describe the datasets used. Section 3 presents the results along with a detailed discussion. Section 4 summarizes the main findings and conclusions of this study.

\section{Data used}
\subsection{Spitzer-IRAC and MIPS data}
To identify young stars in the target region, we used the $Spitzer$ data. We obtained data from the $Spitzer$ Enhanced Imaging Products (SEIP) catalog in the IRAC (3.6 $\mu$m, 4.5 $\mu$m, 5.8 $\mu$m, 8.0 $\mu$m) and MIPS (24 $\mu$m) bands of the $Spitzer$ heritage archive. 
The angular resolution of the IRAC images ranges from $\sim$1.7$^{\prime \prime}$ to 2.0$^{\prime \prime}$, while the MIPS 24 $\mu$m images have a resolution of $\sim$6 $^{\prime \prime}$ \citep{fazio04,rieke04}, enabling the identification and characterization of YSOs over a wide range of evolutionary stages. However, the SEIP data were not available for the entire region (see Fig. \ref{wisemap}) and therefore, to examine the global distribution of the YSOs, we used the WISE data as discussed in the ensuing subsection.

\subsection{Wide-field Infrared Survey Explorer data}
The all-sky coverage, homogeneous sensitivity, and well-matched wavelength range of WISE render it particularly well-suited for investigating large-scale star-forming complexes and identifying spatially distributed YSO populations, including in environments where optical surveys are severely limited or ineffective.
 We used the WISE 3.4 $\mu$m (W1), 4.6 $\mu$m (W2), 12 $\mu$m (W3), and 22 $\mu$m (W4) data with resolutions 6.1$^{\prime \prime}$, 6.4$^{\prime \prime}$, 6.5$^{\prime \prime}$ and 12.0$^{\prime \prime}$, respectively, from the AllWISE Source Catalog \citep{2010AJ....140.1868W}. These data were used to identify and characterize the YSO population within the target region shown in Figure~\ref{wisemap}.

\section{Results and Discussion}
\subsection{Identification of YSOs}
YSOs exhibit characteristic excess emission at infrared wavelengths relative to other astrophysical sources. Consequently, they can be reliably identified using color-color (CC) and color-magnitude (CM) diagrams. The WISE observations encompass the entire spatial extent of the region under investigation, whereas the \textit{Spitzer} data are limited to the area containing the molecular clouds. This complementary coverage allows us to use WISE data to obtain a global view of ongoing star formation across the region, while \textit{Spitzer} observations are used to investigate the spatial distribution and evolutionary stages of YSOs within the molecular clouds.
\subsubsection{YSO identification using Spitzer-IRAC and MIPS data}
Young stellar sources can be identified and classified based on their Mid-Infrared (MIR) colors from the $Spitzer$-IRAC and MIPS observations. However, other contaminants may mimic the colors of the YSOs in the IRAC CC diagrams. These contaminants include star-forming galaxies (SFGs), broad-line active galactic nuclei (AGNs), unresolved knots of shock emissions, and resolved structured PAH emission, which contaminates the photometric apertures of some faint field stars. These sources were rejected using the criteria discussed in \citet{2009ApJS..184...18G}. For example, active SFGs have
strong PAH emission features, which yield very red 5.8 and 8.0 $\mu$m
colors, hence these two bands, alongwith other bands, can be used to filter them out. Similarly, broad-line AGNs have MIR colors that are largely consistent with those of YSOs, so they used the 4.5 and 8.0 $\mu$m CC and CM diagrams to remove them.

We also follow the constraints outlined by \citet{2009ApJS..184...18G} to identify young stellar candidates. The method exploits the infrared excess produced by circumstellar disks and envelopes to distinguish YSOs from field stars. After filtering out contaminants from our data, we selected the YSO candidates using the following criteria:
\begin{equation}
[4.5] - [5.8] > 0.7, ~~~
[3.6] - [4.5] > 0.7
\end{equation}
\begin{equation}
\left.
\begin{aligned}
[4.5] - [8.0] - \sigma_1 > 0.5\\
[3.6] - [5.8] - \sigma_2 > 0.35\\
[3.6] - [5.8] + \sigma_2 <= (0.14/0.04)\times (([4.5] - [8.0] - \sigma_1) - 0.5) + 0.5\\
[3.6] - [4.5] - \sigma_3 > 0.15
\end{aligned}
\right\}
\end{equation}

Where $\sigma_1$ is an error in ([4.5] - [8.0]) color, $\sigma_2$ is an error in ([3.6] - [5.8]), and $\sigma_3$ is an error in ([3.6] - [4.5]) that are calculated using the errors in corresponding magnitudes. The YSO candidates satisfying only equation (2) were classified as Class {\sc ii} YSOs, whereas those fulfilling both equations (1) and (2) were designated as Class {\sc i} sources.

\begin{figure*}
\centering
\includegraphics[width=0.65\textwidth]{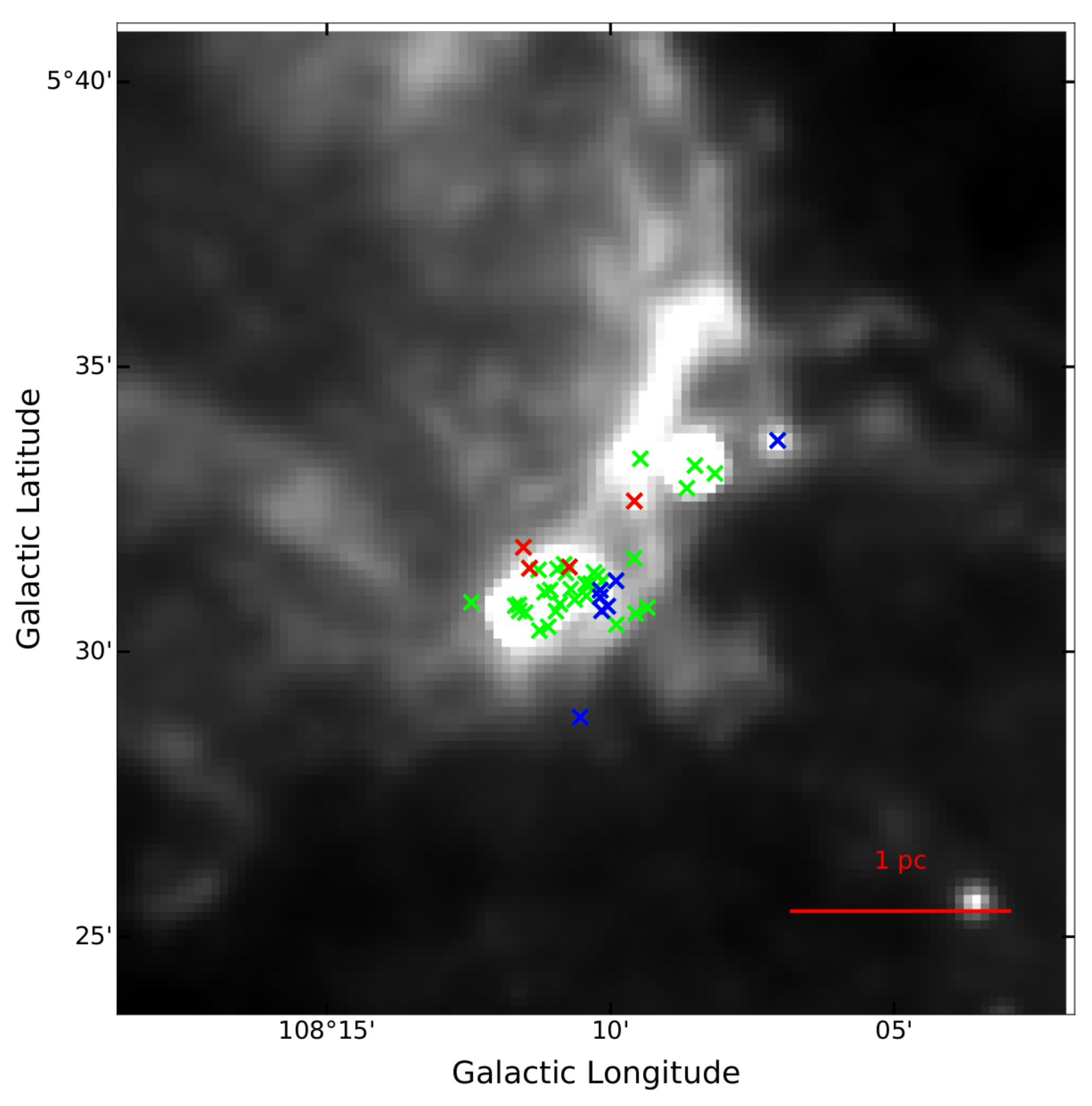}
\includegraphics[width=0.65\textwidth]
{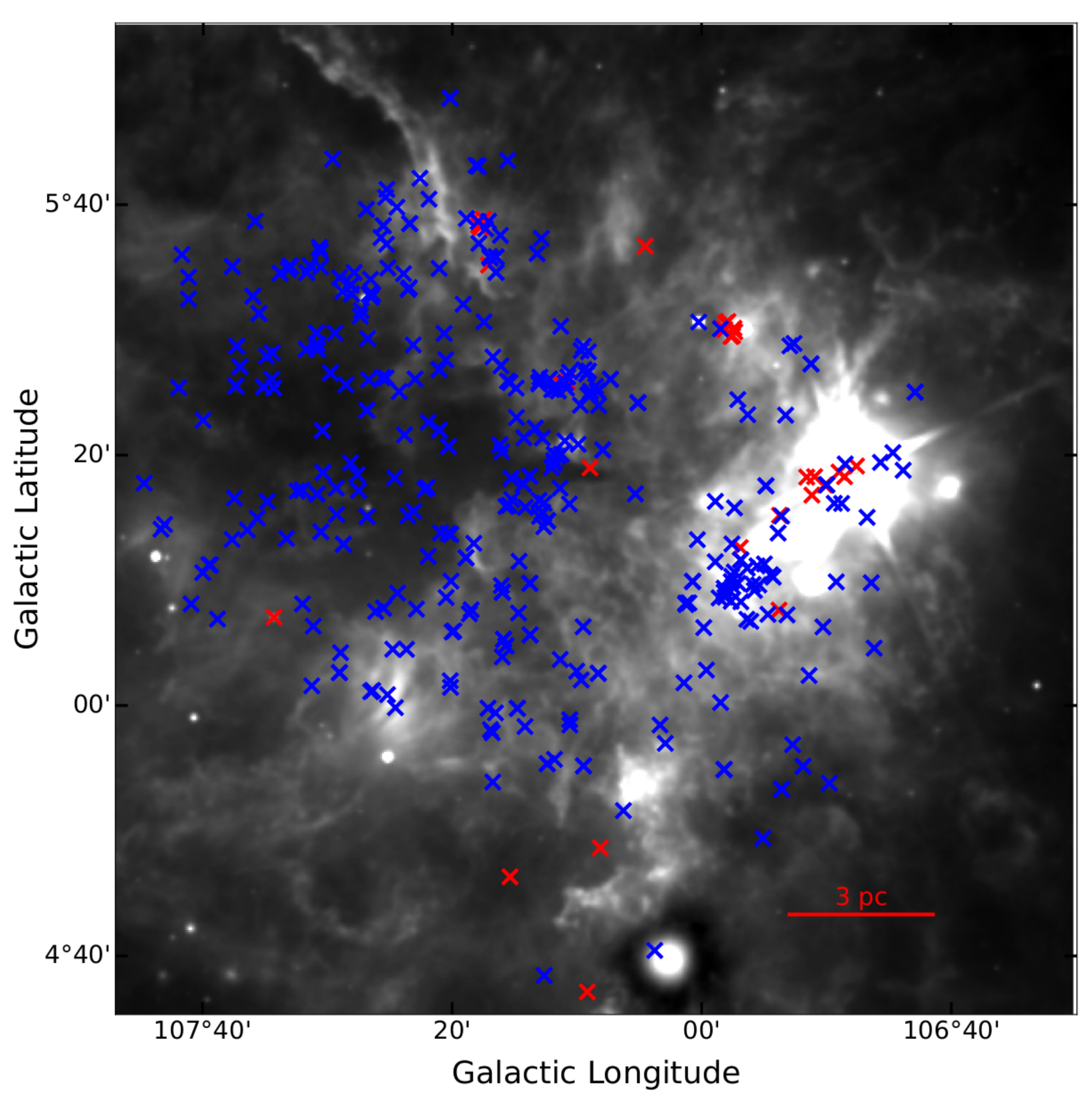}
\begin{minipage}{12cm}
\centering
\begin{minipage}{12cm}
\caption{WISE 22$\mu$m zoomed-in view of the BRC~44 (top panel) and LDN~1204 (bottom panel) regions.  In both panels,  Class~{\sc i} and Class~{\sc ii} YSOs identified in the present work are represented by red and blue cross symbols, respectively.  
The green crosses in top panel represent YSOs identified in our previous study \citep{2025ApJ...994..222C}.} 
\end{minipage}

\label{wisemap_spitz}
\end{minipage}
\end{figure*}

 So, using the magnitudes available in 3.6 $\mu \text{m}$, 4.5 $\mu \text{m}$, 5.8 $\mu \text{m}$, 8 $\mu \text{m}$ and 24 $\mu \text{m}$, we identified around 460 YSOs using the \citet{2009ApJS..184...18G} criterion. The spatial distribution of the identified YSOs is shown in Figure \ref{wisemap_spitz}.

\subsubsection{YSO identification using WISE data}
While $Spitzer$-IRAC and MIPS observations provide highly reliable YSO classification in well-studied regions, the all-sky coverage of the WISE survey enables similar infrared color-based identification over much larger spatial scales, making it particularly valuable for tracing extended and distributed YSO populations beyond the limited fields observed by $Spitzer$.
 The central wavelengths of the WISE photometric bands coincide with the spectral region in which thermal emission from circumstellar disks and envelopes around young stars is prominent. Consequently, \citet{2014ApJ...791..131K} employed WISE observations to distinguish YSOs from other astrophysical sources, such as SFGs and AGNs. They further developed classification criteria to separate YSOs into different evolutionary stages (e.g., Class~{\sc i}, Class~{\sc ii}) by using CC diagrams. In particular, after excluding spurious detections, contaminants, SFGs, and AGNs, they applied the CC selection scheme described below to identify Class~{\sc i} and Class~{\sc ii} sources. By adopting the criteria outlined in \citet{2014ApJ...791..131K}, we removed contaminants from the WISE data of our target region and identified  YSO candidates. After that, we applied the following equations to classify them into the Class~{\sc i} and Class~{\sc ii} sources. 

 \begin{figure*}
\centering
\includegraphics[width=0.75\textwidth]{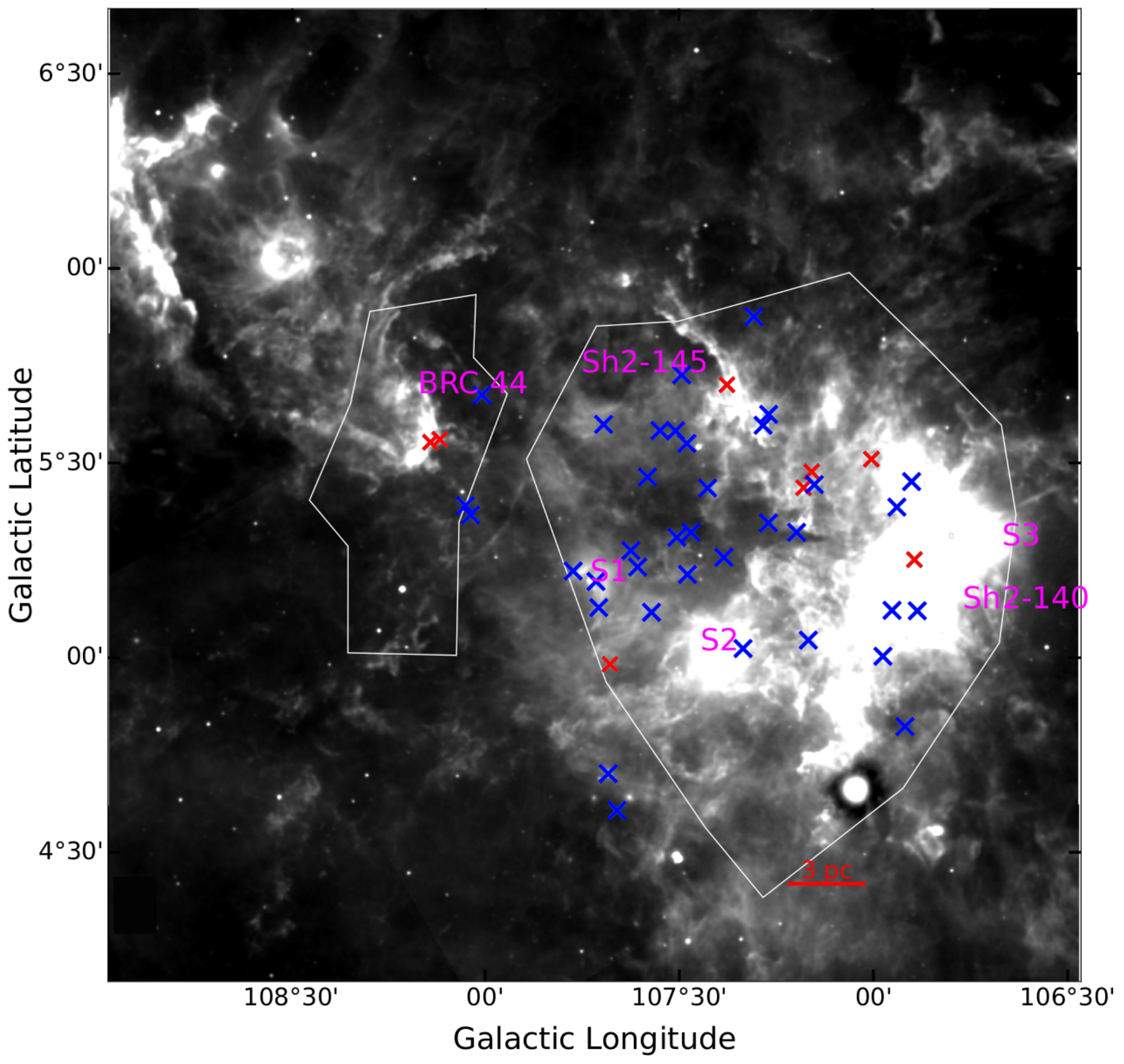}
\centering
\begin{minipage}{12cm}
    
\caption{WISE 22~$\mu$m image of the studied region. The locations of the WISE-identified Class~{\sc i} (red symbols) and Class~{\sc ii} (blue symbols) YSOs are overlaid. The massive stars associated with the H{\sc ii} regions Sh2-145 and Sh2-140, labeled S1, S2, and S3, are also marked, along with the position of BRC~44 \citep{2025ApJ...994..222C}. The white polygon outlines the area covered by the \textit{Spitzer} observations. }
\end{minipage}
\label{wisemap}
\end{figure*}

The YSO candidates were classified as Class {\sc i}, if they satisfy the following equations:
\begin{equation}
\left.
\begin{aligned}
\mathrm{W2} - \mathrm{W3} &> 2.0  \\
\mathrm{W1} - \mathrm{W2} &> -0.42\,(\mathrm{W2} - \mathrm{W3}) + 2.2  \\
\mathrm{W1} - \mathrm{W2} &> 0.46\,(\mathrm{W2} - \mathrm{W3}) - 0.9  \\
\mathrm{W2} - \mathrm{W3} &< 4.5
\end{aligned}
\right\}
\end{equation}

Class {\sc ii} sources satisfy the equations,
\begin{equation}
\left.
\begin{aligned}
\mathrm{w1} - \mathrm{w2} &> 0.25 \\
\mathrm{w1} - \mathrm{w2} &< 0.9\,(\mathrm{w2} - \mathrm{w3}) - 0.25 \\
\mathrm{w1} - \mathrm{w2} &> -1.5\,(\mathrm{w2} - \mathrm{w3}) + 2.1 \\
\mathrm{w1} - \mathrm{w2} &> 0.46\,(\mathrm{w2} - \mathrm{w3}) - 0.9 \\
\mathrm{w2} - \mathrm{w3} &< 4.5
\end{aligned}
\right\}
\end{equation}

\subsection{Spatial distribution of YSOs}
The identified YSOs are overplotted in Fig.   \ref{wisemap}. The two H{\sc ii} regions, Sh2-145 and Sh2-140, are formed by three massive stars, which are also marked. Sh2-145 is formed due to the massive stars HD 213023 (O9V ; `S1' in Fig. \ref{wisemap}) and BD 62 2078 (O7.5V; `S2' in Fig. \ref{wisemap}), whereas Sh2-140 is formed due to HD 211880 (B0.5 V C ; `S3' in Fig. \ref{wisemap} \& \ref{wisemap_spitz}). 
The figure indicates that the spatial distribution of the YSO candidates is non-uniform, with the majority of sources preferentially concentrated toward the north-west of the massive star `S1' rather than toward the south-east. This phenomenon may be due to an enhanced gas density along that line of sight, which subsequently underwent gravitational collapse, leading to the formation of young stellar populations. It can also be noticed that the younger Class {\sc i} YSOs reside inside the cloud compared to Class {\sc ii} candidates, which are distributed both inside and outside the molecular cloud. This sequential evolutionary distribution of candidate YSOs may also point towards the influence of radiative triggering from the massive star(s). 

The H{\sc ii} region Sh2-140, formed by `S3', is smaller compared to Sh2-145, possibly due to the initially very high density of the molecular cloud near `S3' and/or the low-ionizing flux of the massive star. As most of the young stellar candidates are distributed toward the north-west region, it may be due to an initially diffuse and extended distribution of molecular gas cloud toward the south-west periphery of the H{\sc ii} region Sh2-145, as compared to the north-west periphery. 

Fig. \ref{wisemap} clearly shows the `arc-like' emission feature formed by the H{\sc ii} region at one side of the long extended cloud. At one of the ends of the arc (north-east periphery), BRC 44 is located, and at the other end is the dark cloud LDN 1204. This indicates density asymmetry in the surrounding molecular cloud. 
The spatial distribution of the YSOs along with the MIR emission indicates that the massive stars associated with Sh2-145 likely formed near the edge of an extended molecular cloud. The ionizing sources `S1' and `S2', appear to have developed an asymmetric H{\sc ii} region as a consequence of the initially inhomogeneous ambient density distribution. The ionized gas preferentially expanded toward the north, where the surrounding medium was relatively less dense, leading to the formation of the observed arc-like structure and young stars. The compressed molecular material along the ionization front seems to have induced subsequent star formation, particularly in BRC 44, suggesting a feedback-driven triggering mechanism \citep{2025ApJ...994..222C}. The majority of the YSO candidates 
are spatially distributed along the bright rim and within the compressed layer, supporting the scenario of radiation-driven compression and sequential star formation (see Fig. \ref{wisemap_spitz}, top panel). 

In the vicinity of the star `S3', we observe active and ongoing star formation, as evidenced by the presence of numerous Class{\sc i} and Class {\sc ii} YSOs (see Fig. \ref{wisemap_spitz}, bottom panel). The concentration of young sources in this region suggests recent or continuing star-forming activity.

Furthermore, Fig. \ref{wisemap_spitz} (bottom panel) indicates that molecular material appears to be sandwiched between the two adjacent H{\sc ii} regions, Sh2-145 and Sh2-140. This configuration raises the possibility that the interlying molecular cloud may be experiencing compression from both sides, potentially creating favorable conditions for ongoing or future star formation.
However, a detailed molecular line analysis of this region is required to assess 
the kinematic structure and determine whether the cloud dynamics indeed reflect compression driven by the feedback from the 
massive stars. Such an analysis would help constrain whether the 
observed star formation is triggered by external pressure or is part of the intrinsic evolution of the cloud.


\section{Conclusion}

Using the $Spitzer$ and WISE catalog, we identified and classified YSOs across the region encompassing BRC 44 and the dark cloud LDN~1204. The morphology of the cloud and the distribution of YSOs suggest that feedback from nearby massive stars and the associated H{\sc ii} regions has influenced the evolution of the entire complex, including BRC~44 and LDN~1204. Most Class {\sc I} YSOs are concentrated within the cloud, while Class {\sc II} sources are distributed both inside and outside the cloud, indicating a possible age gradient and sequential star formation.
YSOs distributed along the north-west direction, and massive star `S1' may have formed from the same molecular cloud. We are conducting a detailed study using millimeter CO data to understand the cloud dynamics and the nature of star formation across the entire region. 

\begin{acknowledgments}
We express our sincere gratitude to the anonymous referee for constructive comments and suggestions. N.P. and R.C. acknowledge financial support from the Science and Engineering Research Board (SERB) under grant number CRG/2021/005876. This publication makes use of data products obtained with the Spitzer Space Telescope (operated by the Jet Propulsion
Laboratory, California Institute of Technology, under contract
with NASA) and from the Wide-field Infrared Survey Explorer, which is a joint project of the University of California, Los Angeles, and the Jet Propulsion Laboratory/California Institute of Technology, funded by the National Aeronautics and Space Administration. This work was presented as an oral contribution at the BINA-2025 workshop.
\end{acknowledgments}

\begin{furtherinformation}

\begin{orcids}  
\orcid{0009-0007-6397-3220}{Rishi}{C}
\orcid{0000-0002-0151-2361}{Neelam} {Panwar}
\orcid{0000-0001-7881-7748}{Ajay} {Kumar Singh}

\end{orcids}

\begin{authorcontributions}
All the authors have contributed significantly.
\end{authorcontributions}

\begin{conflictsofinterest}
There is no conflict of interest.
\end{conflictsofinterest}

\end{furtherinformation}



%

\bibliographystyle{bullsrsl-en}

\bibliography{extra.bib}

\end{document}